\documentclass{article}

 \usepackage{algorithm}
 \usepackage{algorithmic}
 \usepackage{rotating,subcaption}
\usepackage[preprint]{neurips_2025}

\usepackage[utf8]{inputenc} 
\usepackage[T1]{fontenc}    
\usepackage{hyperref}       
\usepackage{url}            
\usepackage{booktabs}       
\usepackage{amsfonts}       
\usepackage{nicefrac}       
\usepackage{microtype}      
\usepackage{xcolor}         
\usepackage{amsmath}
\usepackage{amssymb}
\usepackage{url}
\usepackage{graphicx}
\usepackage{enumitem}
\usepackage{array}

\title{Adaptive Online Emulation for Accelerating Complex Physical Simulations}

\author{%
Tara P.A. Tahseen$^{1}$\\ \texttt{tara.tahseen.22@ucl.ac.uk} \And Nikolaos Nikolaou$^{1}$ \And Luís F. Simões$^2$ \And Kai Hou Yip$^1$ \And João M. Mendonça$^{3,4}$ \And Ingo P. Waldmann$^1$
\AND $^1${\normalfont Department of Physics \& Astronomy, University College London, London, WC1E 6BT, UK}
\\ $^2${\normalfont ML Analytics, Lisbon, Portugal}
\\ $^3${\normalfont Department of Physics \& Astronomy, University of Southampton, Southampton SO17 1BJ, UK}
\\ $^4${\normalfont School of Ocean and Earth Science, University of Southampton, Southampton, SO14 3ZH, UK}}

\begin{document}

\maketitle

\begin{abstract}
    Complex physical simulations often require trade-offs between model fidelity and computational feasibility. We introduce Adaptive Online Emulation (AOE), which dynamically learns neural network surrogates during simulation execution to accelerate expensive components. Unlike existing methods requiring extensive offline training, AOE uses Online Sequential Extreme Learning Machines (OS-ELMs) to continuously adapt emulators along the actual simulation trajectory. We employ a numerically stable variant of the OS-ELM using cumulative sufficient statistics to avoid matrix inversion instabilities. AOE integrates with time-stepping frameworks through a three-phase strategy balancing data collection, updates, and surrogate usage, while requiring orders of magnitude less training data than conventional surrogate approaches. Demonstrated on a 1D atmospheric model of exoplanet GJ1214b, AOE achieves 11.1× speedup (91\% time reduction) across 200,000 timesteps while maintaining accuracy, potentially making previously intractable high-fidelity time-stepping simulations computationally feasible.
\end{abstract}

\section{Introduction}

Complex physical simulations are fundamental to scientific discovery, but computational costs often force researchers to choose between model complexity and feasibility. Time-stepping simulations in domains like climate modeling \citep{schneider_earth_2017}, molecular dynamics \citep{allen_molecular_2017}, and fluid dynamics \citep{pope_turbulent_2000} are particularly challenging due to their sequential nature and stability requirements \citep{durran_numerical_2010}.

While surrogate modeling approaches replace expensive components with neural network approximations \citep{brunton_data-driven_2019}, existing methods require extensive offline training and struggle when simulations explore previously unseen parameter regions—common in scientific discovery where interesting phenomena occur at boundaries or in rare regimes.

We introduce Adaptive Online Emulation (AOE), which dynamically learns surrogate models during simulation execution using Online Sequential Extreme Learning Machines. AOE addresses two critical limitations: (1) the need for extensive offline training and data generation, and (2) poor generalization across parameter space. By leveraging fast, sample-efficient online learning, AOE continuously refines emulator accuracy along the actual simulation trajectory where needed most.

\section{Methodology: Adaptive Online Emulation}


\subsection{Extreme Learning Machines}


Extreme Learning Machines (ELMs) \citep{schmidt_feedforward_1992, pao_learning_1994, huang_extreme_2004} are single-hidden-layer networks where, given an input matrix $\mathbf{X} \in \mathbb{R}^{N \times d}$ where $N$ is number of samples and $d$ is input dimensionality, a target matrix $\mathbf{Y} \in \mathbb{R}^{N \times m}$ where $m$ is target dimensionality, and a number of hidden layer neurons $H$, input weights $\mathbf{W} \in \mathbb{R}^{d \times H}$ and biases $\mathbf{b} \in \mathbb{R}^{H}$ are randomly fixed, while output weights $\boldsymbol{\beta} \in \mathbb{R}^{H \times m}$ are learned via regularized least squares:

$$\boldsymbol{\beta} = (\mathbf{H}^T\mathbf{H} + \alpha\mathbf{I})^{-1}\mathbf{H}^T\mathbf{Y}$$
where $\mathbf{H} \in \mathbb{R}^{N \times H}$ is the hidden layer output matrix computed as $\mathbf{H} = g(\mathbf{X}\mathbf{W} + \mathbf{1}\mathbf{b}^T)$, $g(\cdot)$ the nonlinear activation function, $\mathbf{1} \in \mathbb{R}^{N}$ a vector of ones for bias broadcasting, $\alpha > 0$ the regularization parameter, and $\mathbf{I} \in \mathbb{R}^{H \times H}$ the identity matrix. This closed-form solution enables rapid training with computational complexity $\mathcal{O}(NH^2 + NHm + H^3)$.

\subsection{Numerically Stable Variant of OS-ELM}

To support our adaptive emulation methodology, we employ a numerically stable variant of OS-ELM \citep{liang_fast_2006} that maintains cumulative sufficient statistics rather than iteratively updating matrix inverses \citep{foster_online_2018, haykin_adaptive_2002}. 

The key insight is that the regularized least squares solution can be expressed using sufficient statistics matrices that decompose additively across data batches:
$$\boldsymbol{\beta}_t = (\mathbf{S}_t^{HH} + \lambda\mathbf{I}_H)^{-1}\mathbf{S}_t^{Hy}$$
where $\mathbf{S}_t^{HH} = \sum_{j=0}^{t} \mathbf{H}_j^T\mathbf{H}_j$ and $\mathbf{S}_t^{Hy} = \sum_{j=0}^{t} \mathbf{H}_j^T\mathbf{Y}_j$.

Upon receiving batch $(\mathbf{X}_t, \mathbf{Y}_t)$, we: (1) compute $\mathbf{H}_t = \sigma(\mathbf{X}_t \mathbf{W})$, (2) accumulate $\mathbf{S}_t^{HH} = \mathbf{S}_{t-1}^{HH} + \mathbf{H}_t^T\mathbf{H}_t$ and $\mathbf{S}_t^{Hy} = \mathbf{S}_{t-1}^{Hy} + \mathbf{H}_t^T\mathbf{Y}_t$, and (3) periodically resolve $\boldsymbol{\beta}_t$ every $T$ updates. 

This approach avoids the numerical instability of traditional OS-ELM's iterative matrix inversions \citep{huang_trends_2015} while providing exact solutions for all accumulated data. The computational complexity is $\mathcal{O}(ldH + lH^2 + lHm + H^3/T + H^2m/T)$ per update.

\subsection{Integration with Time-Stepping Frameworks}

Our framework implements a state machine governing when to use numerical computation, collect data, train, update, or evaluate the surrogate. The simulation operates through three phases:

\textbf{Phase 1: Initialization} ($1 \leq t \leq N_\text{init}$) uses numerical computation to handle initial transients.

\textbf{Phase 2: Training} ($N_\text{init} < t \leq N_\text{init} + N_\text{train}$) collects input-output pairs and performs initial ELM training (at $t=N_\text{init} + N_\text{train}$).

\textbf{Phase 3: Adaptive Execution} ($t>N_\text{init} + N_\text{train}$) alternates in fixed cycles of length $N_\text{cycle} = N_\text{update} + I_\text{update}$, between (1) data collection for $N_\text{update}$ timesteps, (2) an OS-ELM $\boldsymbol{\beta}$ update, and (3) surrogate usage for $I_\text{update}$ timesteps, where $I_\text{update}\gg N_\text{update}$.


The computational complexity per adaptive cycle is $\mathcal{O}(N_\text{update}H^2 + H^3 + I_\text{update}dH)$, where the first term represents statistics accumulation during data collection, the second term represents weight updates, and the third term represents cheap surrogate evaluations. 

\section{Results}

We evaluate AOE on a 1D atmospheric model (the \texttt{OASIS} model; \cite{mendonca_modelling_2020}) of exoplanet GJ1214b over 200,000 timesteps. The simulation models pressure-temperature (p-T) evolution through computationally intensive radiative transfer calculations. The ELM predicts radiative flux profiles that then undergo spatial differentiation to compute atmospheric heating rates, and thus compute the atmospheric p-T profile per timestep. The ELM uses $H=\textbf{1000}$ hidden neurons with $d= 200 \text{ atmospheric layers} \times 3 \text{ physical variables}= \textbf{600}$ input features and $m=201 \text{ atmospheric levels} \times 2 \text{ targets } \times 8 \text{ directions}=\textbf{3216}$ outputs, using a warm-up period of$ N_\text{init}=\textbf{5000}$ and collecting $N_\text{train}=\textbf{5000}$ initial samples followed by periodic $\beta$ updates every $I_\text{update}=\textbf{5100}$ timesteps using $N_\text{update}=\textbf{100}$ update samples. Table \ref{tab:adaptive_performance} includes the timings of components of our AOE framework applied to our test simulation, benchmarked on one NVIDIA V100-16GB GPU.


\begin{figure}
\setkeys{Gin}{width=\linewidth} 
\captionsetup[subfigure]{labelformat=empty} 

    \begin{subfigure}[t]{0.48\linewidth}
    \includegraphics{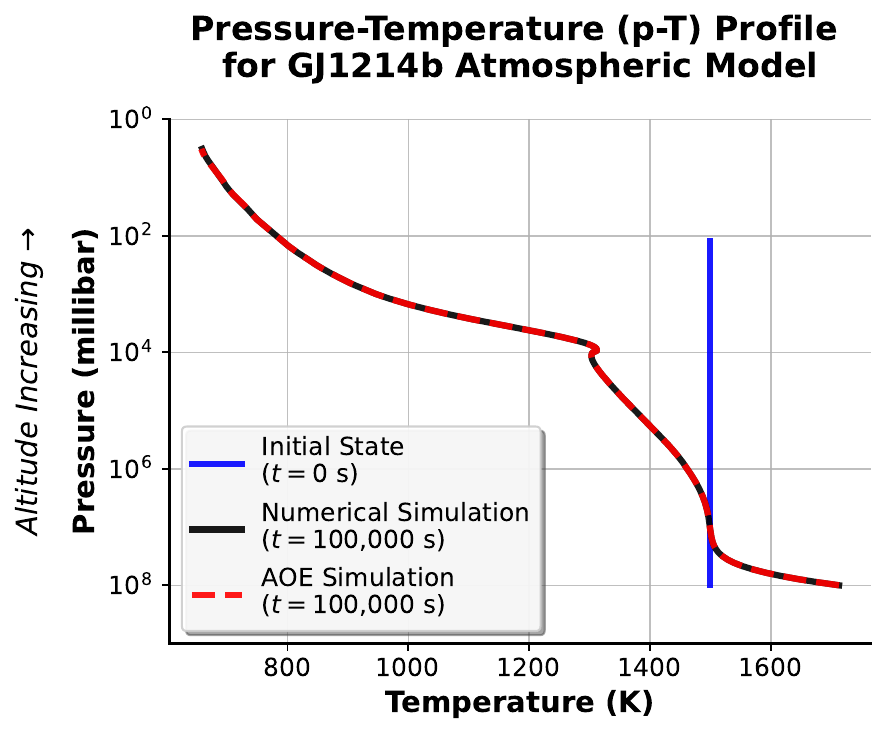}
    \end{subfigure}
    \hfill
    \begin{subfigure}[t]{0.48\linewidth}
    \includegraphics{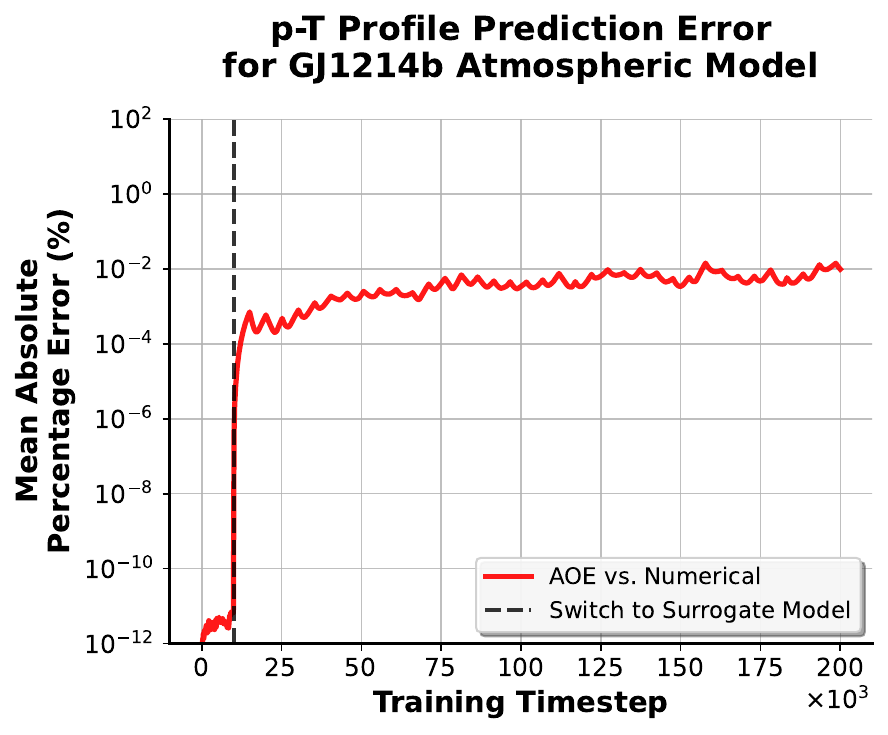}
    \end{subfigure}

\caption{AOE performance on GJ1214b atmospheric simulation. (Left) Pressure-temperature profiles showing strong agreement between numerical (black) and AOE (red dashed) solutions after 200,000 timesteps ($\Delta t = 0.5s$). (Right) Prediction error remains below $\sim0.01$\% throughout the simulation.}
\label{fig:GJ1214b-results}
\end{figure}

\begin{figure}[htbp]
\centering
\begin{subfigure}[t]{0.48\linewidth}
\includegraphics[width=\linewidth]{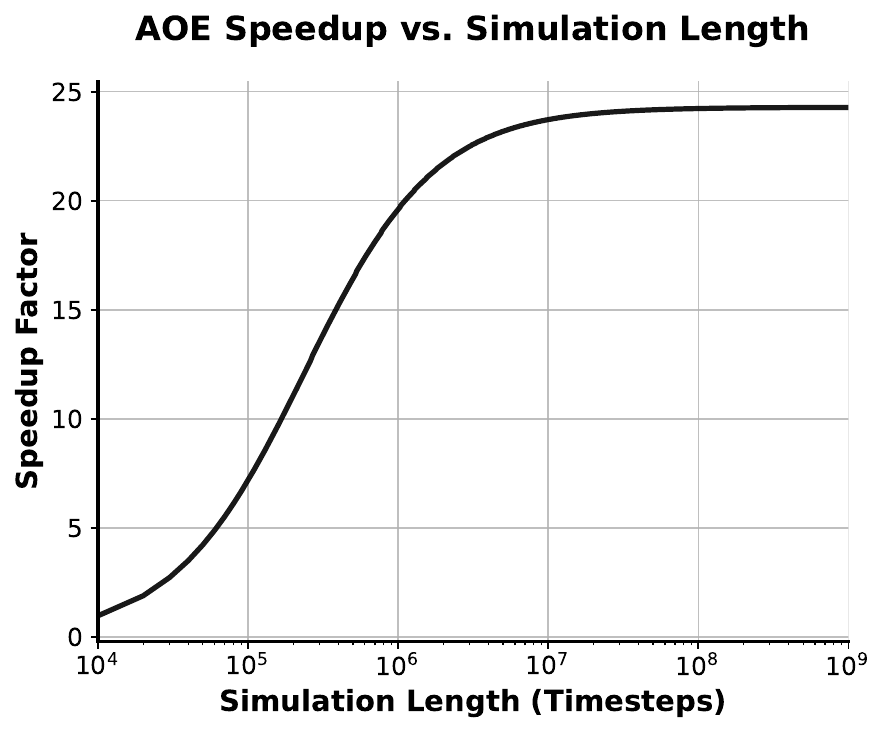}
\end{subfigure}
\caption{Fixed initialization 
costs (warmup + training) are amortized over longer simulations enabling a larger speedup. While this speedup factor is unique to this particular simulation and the chosen AOE hyperparameters ($N_\text{update}, N_\text{cycle}, H)$, the overall trend is simulation-agnostic.}
\label{fig:speedup-vs-timesteps}
\end{figure}


\begin{table}[htbp]
\centering
\caption{Computation time breakdown of Adaptive Online Emulator over 200,000 timesteps benchmarked on one NVIDIA V100-16GB GPU. System overhead includes copying data between arrays and preprocessing and postprocessing input and output data.}
\label{tab:adaptive_performance}
\begin{tabular}{lrrrr}
\toprule
\textbf{Simulation Phase} & \textbf{Duration} & \textbf{Time/Step} & \textbf{Total Time} & \textbf{Speedup} \\
\midrule
\multicolumn{5}{l}{\textit{Initialization Phase}} \\
Warmup (numerical only) & 5,000 steps & 14.14 ms & 70.7 s & 1.0× \\
Data collection (numerical + overhead) & 5,000 steps & 14.21 ms & 71.0 s & 1.0× \\
Initial training & 1 event & 3.09 s & 3.1 s & -- \\
\midrule
\multicolumn{5}{l}{\textit{Emulation Phase}} \\
ML prediction & 186,300 steps & 0.19 ms & 35.4 s & 74.4× \\
System overhead & 186,300 steps & 0.12 ms & 21.6 s & -- \\
\midrule
\multicolumn{5}{l}{\textit{Adaptive Learning}} \\
Retraining data collection & 3,700 steps & 14.21 ms & 52.6 s & 1.0× \\
Model updates & 36 events & 17.8 ms & 0.6 s & -- \\
\midrule
\textbf{Effective emulation} & \textbf{186,300 steps} & \textbf{0.31 ms} & \textbf{57.0 s} & \textbf{46.4×} \\
\textbf{Total simulation} & \textbf{200,000 steps} & \textbf{1.28 ms avg} & \textbf{255.0 s} & \textbf{11.1×} \\
\textbf{Time if all numerical} & \textbf{200,000 steps} & \textbf{14.14 ms} & \textbf{2,828.0 s} & \textbf{1.0×} \\
\midrule
\textbf{Net time savings} & -- & -- & \textbf{2,573.0 s} & \textbf{91\% faster} \\
\bottomrule
\end{tabular}
\end{table}

\section{Discussion}



The AOE framework demonstrates high accuracy in predicting atmospheric pressure-temperature profiles, as shown in Figure \ref{fig:GJ1214b-results}. The emulated simulation closely tracks the numerical reference throughout the 200,000 timestep simulation, with the final atmospheric state (red dashed line) nearly indistinguishable from the pure numerical result (black line). Mean absolute percentage errors remain consistently around $0.01$\% for the majority of the simulation. Notably, this high accuracy is achieved despite the challenging nature of the prediction task: the ELM predicts radiative flux profiles that then undergo spatial differentiation to compute heating rates, meaning small prediction errors could potentially be amplified in the temperature evolution. This level of accuracy, maintained over an extended simulation period, validates the viability of online emulation for scientific computing applications that require large numbers of time steps.

The AOE framework achieves substantial computational acceleration, reducing simulation time from 2,828 seconds (pure numerical) to 255 seconds (AOE-enabled), representing a $11.1\times$ speedup. This performance improvement scales favorably with simulation length as shown in Figure \ref{fig:speedup-vs-timesteps}: while initial phases require numerical computation for data collection and training, the majority of timesteps benefit from fast surrogate evaluation. The speedup increases asymptotically with longer simulations as the fixed initialization costs are amortized over more emulated timesteps.

To contextualize this acceleration, the original numerical computation involves a complex CUDA kernel that executes across multiple dimensions: spectral channels, spatial layers, and 8 directional sampling points. For each combination, the kernel performs multi-dimensional interpolations across 4D lookup tables (pressure × temperature × species × wavelength), computes scattering coefficients, applies transport calculations through 200 atmospheric layers, and solves coupled differential equations for upward/downward propagation. This multi-dimensional computation with nested loops, table interpolations, and iterative solvers requires 14.14ms per timestep. In contrast, surrogate evaluation involves a single forward pass through the ELM network requiring only 0.31 ms per timestep (including system overhead of copying and scaling data)—a $46\times$ acceleration per timestep where emulation is used.

The overall simulation acceleration of $11.1\times$ represents a conservative baseline using fixed hyperparameters rather than optimized control policies. The update frequency $N_\text{cycle}$ and update batch size $N_\text{update}$ directly control the accuracy-efficiency trade-off, and intelligent optimization of these parameters could yield additional speedup while maintaining high accuracy.

\section{Future Work}



The current AOE framework has several key areas for improvement that motivate future research:


\begin{enumerate}

    \item \textbf{Adaptive control policies:} Replace fixed hyperparameters  $N_{\text{update}}$ and  $N_{\text{cycle}}$ with data-driven policies that adjust update frequency based on observable metrics such as recent prediction variance, rate of state change, or time since last update.
    \item \textbf{Active Learning:} Implement methods to intelligently sample input space for training and updates to boost emulator performance.
    \item \textbf{Operational quality control:} Implement real-time surrogate validation through uncertainty quantification (using prediction confidence bounds or ensemble variance estimates).

\end{enumerate}
These developments could significantly enhance the robustness of the AOE framework while enabling greater computational acceleration through intelligent adaptation. The demonstrated effectiveness of AOE on atmospheric modeling suggests potentially broad applicability to computationally intensive scientific domains where expensive physics calculations dominate runtime. By making previously prohibitive simulations tractable while maintaining the accuracy essential for reliable scientific insights, Adaptive Online Emulation has substantial potential to accelerate scientific discovery across physics-based modeling applications.

\section{Code Availability}
The C++/CUDA implementation of our Adaptive Online Emulator framework will be made publicly available on GitHub on August 29th 2025.

\bibliographystyle{plainnat}
\bibliography{references}

\end{document}